\documentclass[reprint,aps,prl,superscriptaddress,lengthcheck]{revtex4-2}
\usepackage{epsfig}
\usepackage{latexsym}
\usepackage{xspace}
\usepackage[colorlinks=true,linktocpage=true,linkcolor=blue,citecolor=blue,allcolors=blue]{hyperref}
\usepackage[utf8]{inputenc}
\usepackage{indentfirst}
\usepackage{enumerate}
\usepackage{color}

\usepackage[caption=false,position=top]{subfig}

\usepackage{amsmath}
\usepackage{amssymb}
\usepackage[english]{babel}
\usepackage{url}

\usepackage{pdfpages}
\makeatletter
\AtBeginDocument{\let\LS@rot\@undefined}
\makeatother

\newcommand{\eq}[1]{\begin{align} #1 \end{align}}
\newcommand{\mean}[1]{\langle #1 \rangle}

\newcommand{\sNN}{\sqrt{s_{\rm NN}}}

\newcommand{\rr}{{\bf r}}

\begin{document}

\title{
Indications for freeze-out of charge fluctuations in the quark-gluon plasma at the LHC
}

\author{Jonathan~Parra}
\affiliation{Physics Department, University of Houston, Box 351550, Houston, TX 77204, USA}

\author{Roman~Poberezhniuk}
\affiliation{Physics Department, University of Houston, Box 351550, Houston, TX 77204, USA}
\affiliation{Bogolyubov Institute for Theoretical Physics, 03680 Kyiv, Ukraine}

\author{Volker~Koch}
\affiliation{Nuclear Science Division, Lawrence Berkeley National Laboratory, 1 Cyclotron Road, Berkeley, CA 94720, USA}

\author{Claudia~Ratti}
\affiliation{Physics Department, University of Houston, Box 351550, Houston, TX 77204, USA}

\author{Volodymyr~Vovchenko}
\affiliation{Physics Department, University of Houston, Box 351550, Houston, TX 77204, USA}

\begin{abstract}

The D-measure of net-charge fluctuations quantifies the variance of net charge in strongly interacting matter. 
It was introduced over 20 years ago as a potential signal of quark-gluon plasma (QGP) in heavy-ion collisions, 
where it is expected to be suppressed due to the fractional electric charges of quarks.
Measurements have been performed at RHIC and LHC, but the conclusion has been elusive in the absence of quantitative calculations for both scenarios. 
We address this issue by employing a recently developed formalism of density correlations and incorporate resonance decays, local charge conservation, and experimental kinematic cuts. 
We find that the hadron gas scenario is in fair agreement with the ALICE data for $\sNN = 2.76$~TeV Pb–Pb collisions only when a very short rapidity range of local charge conservation is enforced, while the QGP scenario is in excellent agreement with experimental data and largely insensitive to the range of local charge conservation.
A Bayesian analysis of the data utilizing different priors yields moderate evidence for the freeze-out of charge fluctuations in the QGP phase relative to hadron gas.
The upcoming high-fidelity measurements from LHC Run 2 will serve as a precision test of the two scenarios.

\end{abstract}


\maketitle

\emph{Introduction.---}
Heavy-ion collisions at ultrarelativistic energies provide a unique environment for studying matter under extreme conditions, where the formation of the quark–gluon plasma (QGP) is expected to occur.
From first-principles lattice QCD calculations, it is now understood that the deconfinement transition is a smooth crossover at LHC energies~\cite{Aoki:2006we} as well as across a broad range of RHIC energies~\cite{HotQCD:2018pds,Borsanyi:2020fev}.
Numerous measurements at the LHC and RHIC have provided evidence for the creation of a hot and strongly interacting medium consistent with QGP formation~\cite{Harris:1996zx,Bass:1998vz,BRAHMS:2004adc,PHOBOS:2004zne,STAR:2005gfr,PHENIX:2004vcz,ALICE:2022wpn}~(see Ref.~\cite{Harris:2023tti} for a recent overview).
Event-by-event fluctuations can serve as a distinct QGP signal due to their sensitivity to the relevant degrees of freedom in the medium~\cite{Asakawa:2015ybt,Luo:2017faz,Bzdak:2019pkr}.  
In particular, the D-measure of net-charge fluctuations~\cite{Asakawa:2000wh,Jeon:2000wg},
\eq{\label{eq:D}
D=4\frac{\kappa_2[Q]}{\mean{N_{\rm ch}}},
}
was introduced to quantify the anticipated suppression of fluctuations arising from the fractional charges of quarks in a deconfined phase~\cite{Jeon:2000wg}. 
Here, $Q = N_+ - N_-$ is the measured net charge~\footnote{We neglect multi-charged hadrons in the final state such as light nuclei.},
$N_{\pm}$ are the measured multiplicities of positively and negatively charged particles, $N_{\rm ch}=N_+ + N_-$ is the total charged particle multiplicity,
$\mean{...}$ denotes event-by-event averaging, and 
$\kappa_2[Q]=\mean{(Q^{\rm acc})^2}-\mean{Q^{\rm acc}}^2$ is the variance of net-charge distribution.
As the net charge $Q$ is conserved in full acceptance, the D-measure~\eqref{eq:D} is only meaningful within a finite acceptance, which, however, must not be too small, as fluctuations then approach the Poisson limit~\cite{Koch:2008ia}.
 
Since quarks carry fractional charges, the D-measure is expected to be significantly smaller in the QGP.
For a pion gas in the grand-canonical ensemble~(GCE), one has $D_{\pi}\approx 4$ reflecting their unit charge and Poisson statistics, while a smaller value of $D_{\rm HG} \approx 3$~\cite{Jeon:2000wg} is expected in a hadron gas~(HG) due to resonance decays.
Obtaining the estimate for QGP is less straightforward, mainly because $\mean{N_{\rm ch}}$ is not defined in the QGP phase.
The estimates based on entropy conservation and values of charge susceptibility in QGP yield $D_{\rm QGP} \approx 1$~\cite{Jeon:2000wg}, about three times smaller than the hadron gas value.

The global estimates mentioned above correspond to the GCE limit. 
However, the measurements in heavy-ion collisions are subject to kinematic cuts, and they are affected by the exact (local) conservation of electric charge~\cite{Bleicher:2000ek,Sakaida:2014pya,Vovchenko:2020tsr} and diffusive dynamics~\cite{Shuryak:2000pd,Aziz:2004qu}.
For this reason, the interpretation of experimental measurements of the D-measure by STAR~\cite{STAR:2003oku,STAR:2008szd} and ALICE~\cite{ALICE:2012xnj}, which tend to lie in between the naive global estimates for HG and QGP, has been inconclusive.
Kinematic cuts and global charge conservation have been considered in hadronic gas calculations~\cite{Bleicher:2000ek,Vovchenko:2020kwg}, but this has not been sufficient to describe the data.
The D-measure has been calculated using microscopic transport models at LHC~\cite{Mishra:2017bdq} and RHIC~\cite{Zhang:2002dy,Singh:2019skh} energies, and its connection to charge balance functions~\cite{Bass:2000az,Pratt:2021xvg} was explored~\cite{Ling:2013ksb}. 
However, drawing definitive conclusions remains challenging due to the lack of quantitative modeling of HG and QGP scenarios on an equal footing.

In this Letter, we present quantitative calculations for both the HG and QGP scenarios for Pb–Pb collisions at LHC energies.
First, we revisit the estimates for the normalized charge fluctuation $\omega = \kappa_2[Q]/\mean{N_{\rm ch}^{\rm prim}}$ at hadronization, which is expressed in terms of charge susceptibility $\chi_2^Q$ and entropy density $s$ at the stage where charge fluctuations freeze out and distinguish HG and QGP scenarios.
We then incorporate the effects of resonance decays, local charge conservation, and kinematic cuts by using the 2-point charge density correlator within a recently developed framework~\cite{Vovchenko:2024pvk}, leading to sizable quantitative changes compared to the previous global estimates~\footnote{We note that the global estimates of \cite{Jeon:2000wg} were rather crude as they did not include effects of resonance decays for the QGP scenario and had no $p_T$ cuts. Therefore, they do not provide a proper reference for a quantitative comparison.}.
We perform a Bayesian analysis of experimental data of the ALICE Collaboration for Pb–Pb collisions at $\sNN = 2.76$~TeV and obtain moderate evidence for freeze-out in the QGP.
The upcoming high-statistics data from LHC Run 2 can provide a precision test of these scenarios.

\emph{Fluctuations at hadronization.---} 
The starting point of our analysis is the hadronization stage.
We study the following scenario.
The mean charged multiplicity $\mean{N_{\rm ch}^{\rm prim}}$ is in chemical equilibrium and determined by the Hadron Resonance Gas (HRG) model, carrying no memory of QGP.
However, the variance of a conserved net-charge, $\kappa_2[Q]$, may only evolve through diffusion.
The diffusion process may be too slow to maintain equilibrium and thus the net-charge variance may decouple and freeze-out earlier than the mean multiplicities, possibly in the QGP.

We quantify charge fluctuations at hadronization through
\eq{
\omega = \frac{\kappa_2[Q]}{\mean{N_{\rm ch}^{\rm prim}}}.
}
Here $\kappa_2[Q]$ is the charge variance and $\mean{N_{\rm ch}^{\rm prim}}$ the multiplicity of charged hadrons at hadronization, i.e. before resonance decays. 
We assume the GCE for the time being.
If fluctuations freeze out in the hadron gas, one expects the Poisson baseline of $\omega_{\rm HG} \approx 1$.
A quantitative calculation within the HRG model~\cite{Vovchenko:2019pjl} gives a slightly larger value, $\omega_{\rm HG} \approx 1.1$ at $T = 155-160$~MeV, with the enhancement attributed to the Bose-Einstein statistics for pions and the presence of multi-charged hadrons.

Estimating $\omega$ for the QGP phase is more tricky, as $\mean{N_{\rm ch}^{\rm prim}}$ is not well defined outside the hadronic phase.
One thus typically resorts to using the entropy density $s$ as a proxy~\cite{Bass:1998vz,Asakawa:2000wh,Jeon:2000wg}, 
utilizing the approximately isentropic evolution between the freeze-out of charge fluctuations and hadronization.
In the grand-canonical ensemble, the charge susceptibility is a measure of net-charge variance per unit volume, thus $\kappa_2[Q] = V \chi_2^Q$.
Dividing and multiplying $\omega$ by the entropy $S$ and then by the final charged multiplicity $\mean{N_{\rm ch}}$ one can write
\eq{\label{eq:omega}
\omega 
& = \frac{\kappa_2[Q]}{\mean{N_{\rm ch}^{\rm prim}}} 
= \frac{V \chi_2^Q}{S} \frac{S}{\mean{N_{\rm ch}^{\rm prim}}}  = \frac{\chi_2^Q}{s} \frac{S}{\mean{N_{\rm ch}}} \frac{\mean{N_{\rm ch}}}{\mean{N_{\rm ch}^{\rm prim}}}.
}

Here the charge susceptibility $\chi_2^Q$ and entropy density $s$ are given by the equation of state, such as the free~(Stefan-Boltzmann) QGP limit or lattice QCD~\footnote{In lattice QCD one commonly uses dimensionless charge susceptibility $\tilde \chi_2^Q$, which differs from the one used here by factor $T^3$, namely $\tilde \chi_2^Q = \chi_2^Q / T^3$.}. 
$S/\mean{N_{\rm ch}}$ is the entropy per charged hadron in the final state.
While earlier estimates for $S/\mean{N_{\rm ch}}$ had varied significantly,
recent studies at LHC energies give the value $S/\mean{N_{\rm ch}} = 6.7 \pm 0.8$~\cite{Hanus:2019fnc} which we 
employ here.
The last term is the ratio $\mean{N_{\rm ch}}/\mean{N_{\rm ch}^{\rm prim}}$ of the final and primordial charged multiplicities, which is enhanced relative to unity by resonance decays.
A thermal model estimate at $T = 155$~MeV gives $\gamma_Q = \mean{N_{\rm ch}}/\mean{N_{\rm ch}^{\rm prim}} \approx 1.67$~\cite{Vovchenko:2019pjl}.
For a free gas of massless $u,d,s$-quarks and gluons one has
$(\chi_2^{Q})_{\rm QGP} / T^3 = 2/3$ and $s_{\rm QGP}/ T^3 = 19\pi^2 / 9$,
therefore
\eq{
\omega_{\rm QGP} = 0.36 \pm 0.04.
}
The uncertainty here comes solely from $S/\mean{N_{\rm ch}}$.

\begin{figure}[t]
  \centering
  \includegraphics[width=.49\textwidth]{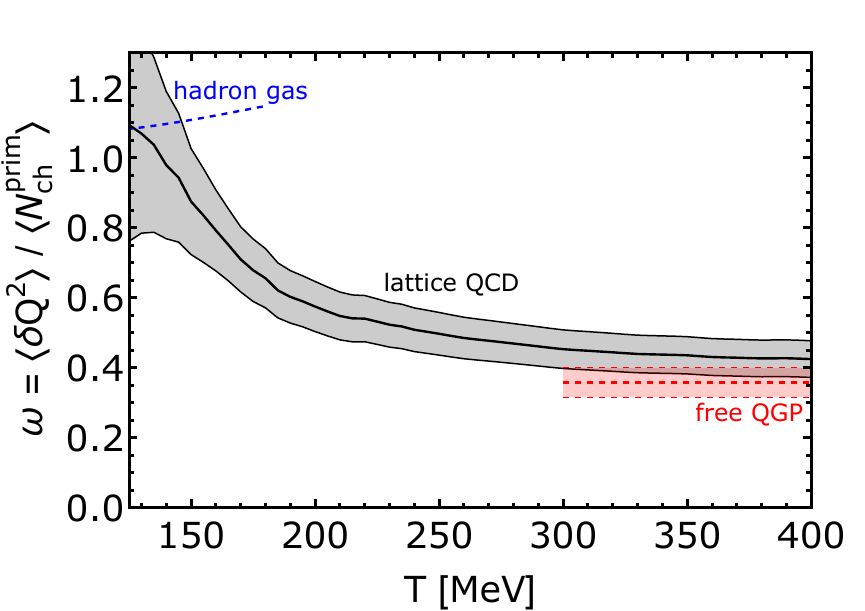}
  \caption{
  Temperature dependence of the normalized charge variance $\omega = \kappa_2[Q]/\mean{N_{\rm ch}^{\rm prim}}$ computed through Eq.~\eqref{eq:omega} using lattice QCD data for $\chi_2^Q$ and $s/T^3$~(gray band), as well as the hadron resonance gas~(dashed blue line) and free massless QGP~(dashed red line) limits.
  The error bands reflect the uncertainties in lattice QCD data~\cite{Borsanyi:2011sw,Borsanyi:2013bia} and $S/\mean{N_{\rm ch}}$~\cite{Hanus:2019fnc}.
  }
  \label{fig:wLQCD}
\end{figure}

Equation~\eqref{eq:omega} defines the value of $\omega$ in a more general case 
beyond just HG and QGP limits.
In particular, one can utilize lattice QCD data for $\chi_2^Q$ and $s/T^3$ to evaluate
$\omega$ 
for a freeze-out of charge fluctuations at a particular temperature.
The corresponding temperature dependence of $\omega$ is depicted in Fig.~\ref{fig:wLQCD}, utilizing the lattice QCD data of the Wuppertal–Budapest collaboration~\cite{Borsanyi:2011sw,Borsanyi:2013bia}.
The lattice-based result for $\omega$ is consistent within errors with a smooth evolution between the hadron gas regime at $T \lesssim 130-160$~MeV~($\omega \simeq 1.1$) and the QGP regime at higher temperature, with $\omega \approx 0.40-0.45$ at $T \geq 300$~MeV being close to the Stefan-Boltzmann limit of $\omega_{\rm QGP} \simeq 0.36$.
If $\omega$ can be reliably extracted from the data, one could map it to the temperature of freeze-out of charge fluctuations.

\emph{Including experimental effects.---}
While the difference between the HG and QGP scenarios is evident in the resulting values of $\omega$, this quantity cannot be directly compared to experimental measurements. 
Such a comparison would be distorted by several experimental effects:
\begin{itemize}
\item Resonance decays in the hadronic phase.
Decays of neutral resonances, such as $\rho_0 \to \pi^+ \pi^-$, create correlated pairs of charges. 
Although such decays do not change the net charge number, they do increase the charged multiplicity in Eq.~\eqref{eq:D} by a substantial factor ($\gamma_Q \simeq 1.67$).

\item Exact charge conservation. $\omega$ in Eq.~\eqref{eq:omega} quantifies fluctuations in the grand-canonical limit.
Experimental measurements are affected by canonical effects, where the total net charge is conserved and all produced charges are balanced by the corresponding opposite charges in the fireball. This leads to suppression of fluctuations relative to the naive grand-canonical expectation~\cite{Bleicher:2000ek,Sakaida:2014pya,Vovchenko:2020tsr}.

\item Kinematical cuts. A measure such as $\omega$ describes fluctuations in coordinate space, whereas the experimental measurements are performed in momentum space. The relation between the two is non-trivial due to collective flow in heavy-ion collisions.
The presence of momentum cuts dilutes the strength of two-particle correlations measured by the detector.

\end{itemize}

Here, we incorporate these effects by employing a differential density-density correlator in spatial rapidity.
For a longitudinally boost invariant thermal system with local charge conservation this quantity reads~\cite{Vovchenko:2024pvk}:
\eq{\label{eq:C2Q}
\mathcal{C}^Q_2(\eta_1,\eta_2) = \chi_2^Q \left[ \delta(\eta_1 - \eta_2) - \frac{\varkappa(\eta_1,\eta_2)}{2 \eta_{\rm max}} \right].
}
Here $\chi_2^Q$ is the grand-canonical charge susceptibility, $\eta_{\rm max}$ is the spatial rapidity cut-off, and $\varkappa(\eta_1,\eta_2)$ is the local charge conservation kernel, which we model by a Gaussian~\cite{Vovchenko:2024pvk}.
The first term in Eq.~\eqref{eq:C2Q} contains contributions from self-correlation and local two-particle correlations. 
The relative sizes of these two contributions are sensitive to the value of $\omega$ and resonance decays ($\gamma_Q$).
The second term describes non-local two-particle correlations due to charge conservation.
Kinematic acceptance cuts are introduced through momentum acceptance probabilities $p(\eta)$.

Integrating the density-density correlator 
over the spatial rapidities produces the variance $\kappa_2[Q^{\rm acc}]$ of the net charge in a given momentum acceptance.
We present the technical details for this calculation, as well as the relevant parameters for Pb-Pb collisions at 2.76 TeV, in the End Matter.
The resulting expression for the D-measure inside momentum acceptance, $D = 4 \kappa_2[Q^{\rm acc}] / \mean{N_{\rm ch}^{\rm acc}}$, reads
\eq{\label{eq:Dmeasure:main}
D = 4 \left\{ 1 - \left( 1- \frac{\omega}{\gamma_Q} \right) \frac{\mean{p^2(\eta)}}{\mean{p(\eta)}} - \frac{\omega}{\gamma_Q} \frac{\mean{p(\eta_1) p(\eta_2)}_\varkappa}{\mean{p(\eta)}} \right\}.
}

Equation~\eqref{eq:Dmeasure:main} is the central result of this Letter. 
It quantifies the contributions of the self-correlation~(first term), local~(second term) and non-local~(third term) two-particle correlations to the final result, arising from the simultaneous effects of the possible early freeze-out of fluctuations~($\omega$), resonance decays~($\gamma_Q$), local charge conservation~($\varkappa$), collective flow and kinematic cuts~[$p(\eta)$].
Here $\mean{p(\eta)}$ and $\mean{p^2(\eta)}$ are average single-particle and pair acceptances.
The quantity $\mean{p(\eta_1) p(\eta_2)} = \frac{1}{4 \eta_{\rm max}^2} \iint d \eta_1
d \eta_2  p(\eta_1) p(\eta_2) \varkappa(\eta_1,\eta_2)$ encodes the effect of local charge conservation.

As mentioned above, local charge conservation kernel is modeled by a Gaussian,  $\varkappa (\eta_1,\eta_2) \propto \exp \left[-\frac{(\eta_1-\eta_2)^2}{2 \sigma_{y}^2}\right]$, where $\sigma_y$ quantifies the range of local charge conservation in spatial rapidity.
For convenience, we will use the size of the effective conservation volume $V_c$ relative to the total volume $V_{\rm tot}$ to quantify local charge conservation. 
This ratio is given by~\cite{Vovchenko:2020tsr}
\eq{
\frac{V_c}{V_{\rm tot}} = \sqrt{\frac{\pi}{2}} \frac{\sigma_y}{\eta_{\rm max}} 
{\rm erf } \left(\frac{\eta_{\rm max}}{\sqrt{2} \sigma_{y}}\right).
}
Note that $V_c \to V_{\rm tot}$ as $\sigma_y \to \infty$, as expected for the global charge conservation limit. We treat the ratio $V_c / V_{\rm tot}$ as a free parameter in this study.

\emph{Comparison with experimental data.---}
The ALICE Collaboration reported measurements of the D-measure in Pb–Pb collisions at $\sNN = 2.76$~TeV~\cite{ALICE:2012xnj}.
The measurements were performed in a broad transverse momentum window, $0.2 < p_T < 5.0$~GeV/$c$, as a function of the pseudorapidity cut $|\tilde \eta| < \tilde \eta_{\rm cut}$~(up to $\tilde \eta_{\rm cut} = 0.8$).
We take $\gamma_Q = 1.67$ and contrast hadron gas~($\omega_{\rm HG}=1.1$) and QGP~($\omega_{\rm QGP} = 0.36$) scenarios.
To elucidate the effect of local charge conservation, we performed the calculations for global charge conservation~($V_c = V_{\rm tot}$), and two values of local charge conservation, $V_c = 0.2 V_{\rm tot}$~($\sigma_y = 0.78)$ and $V_c = 0.31V_{\rm tot}$~($\sigma_y = 1.20$)
which are motivated by analyzing local baryon conservation in Ref.~\cite{Vovchenko:2024pvk}.
We use the blast-wave model~\cite{Schnedermann:1993ws} to evaluate acceptance probabilities $p(\eta)$, with the parameters taken from~\cite{ALICE:2013mez}. 
We compute $p(\eta)$ as a weighted average of acceptance probabilities for pions, kaons, and protons, based on their relative measured yields, with the details available in the supplemental material~\cite{SM}.
Finally, we note that the measurements in~\cite{ALICE:2012xnj} are corrected for global charge conservation, namely $D^{\rm corr} = (D' + D'')/2$, where $D'  = D +  4\mean{p(\eta)}$ and $D''  = \frac{D}{1 - \mean{p(\eta)}}$~\footnote{In the notation of \cite{ALICE:2012xnj}, the factor $\mean{p(\eta)}$ reads as $\mean{p(\eta)} = \mean{N_{\rm ch}} / \mean{N_{\rm ch}^{\rm tot}}$, where $\mean{N_{\rm ch}}$ and $\mean{N_{\rm ch}^{\rm tot}}$ are charged multiplicities of accepted and all particles, respectively.}.
We perform the same correction for comparisons with the data.
The value of $\mean{p(\eta)}$ varies between $0.02$ and $0.13$ as a function of $\tilde \eta_{\rm cut}$ and the correction size is consistent with the one used by ALICE~\cite{ALICE:2012xnj}.

\begin{figure}[t!]
    \centering
    \includegraphics[width=1.0\linewidth]{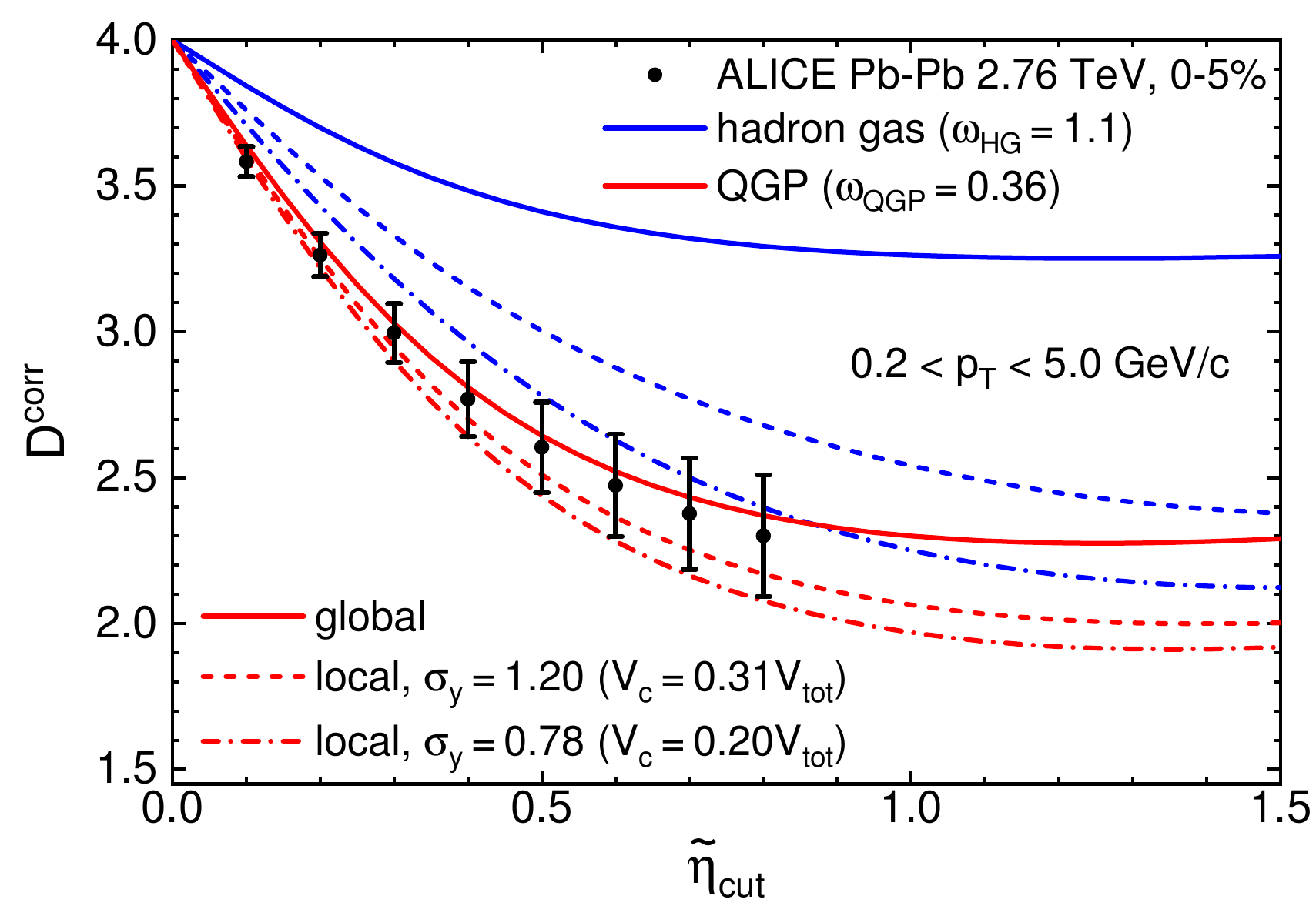}
    \caption{
    The corrected D-measure as a function of pseudorapidity cut $\tilde \eta_{\rm cut}$ in central Pb–Pb collisions at $\sNN = 2.76$~TeV calculated via Eq.~(\ref{eq:Dmeasure}) for hadron gas~(blue lines) and QGP~(red lines) scenarios with global~(solid lines) and local~(styled lines) charge conservation.
    The symbols depict the experimental data of the ALICE Collaboration~\cite{ALICE:2012xnj}.
    }
    \label{fig:DPlotRun1}
\end{figure}

Figure~\ref{fig:DPlotRun1} compares our model calculations for $D^{\rm corr}$ with the experimental data~\cite{ALICE:2012xnj} for $0-5\%$ central collisions.
The hadron gas scenario with only global charge conservation fails to describe the data~(solid blue line). The result is consistent with prior studies~\cite{Vovchenko:2020kwg}.
A better agreement is obtained when significant local conservation is enforced, for instance $\sigma_y = 0.78$ (dash-dotted blue line).
In contrast, the QGP scenario exhibits a much weaker sensitivity to local charge conservation, with fair agreement with experimental data observed for almost any value of $V_c$.
This weak sensitivity is attributed to the fact that the local conservation term~[Eq.~\eqref{eq:Dmeasure}] is proportional to $\omega$, which is more than three times smaller in QGP relative to HG.
We also applied our formalism to other centralities, where we used the corresponding blast-wave parameters. 
Our calculations reproduce the mild centrality dependence observed by ALICE~\cite{ALICE:2012xnj}.

\emph{Bayesian analysis.---}
Our result generally indicates the preference for the QGP scenario, although the HG scenario may also be feasible for a sufficiently local range of charge conservation.
Thus, the conclusion may be sensitive to the prior assumption on the range of local charge conservation.
In addition to the two extreme scenarios considered above, it is also feasible to explore the freeze-out of fluctuations at an intermediate stage between QGP and HG. 
To further investigate this,
we perform a Bayesian analysis of experimental data in the $(\omega,V_c/V_{\rm tot})$ parameter space, utilizing the Gaussian likelihood function.
The experimental uncertainties for different data points are expected to be correlated, given that the data points correspond to overlapping acceptances.
We thus only include the smallest $\tilde \eta_{\rm cut} = 0.1$ and largest $\tilde \eta_{\rm cut} = 0.8$ acceptance data points into the likelihood, for which we can neglect the correlation.

The parameter $\omega$ is varied in the range $\omega \in [0,1.2]$.
In this way, one covers both the HG and QGP scenarios, as well as intermediate values and even an extreme case of vanishing local fluctuations~($\omega \to 0$). We adopt a uniform prior throughout, $\omega \sim U(0,1.2)$, which attributes equal prior probability for HG and QGP.
The conservation volume is varied in the range $V_c/V_{\rm tot} \in [0, 1]$, covering both the highly local and fully global charge conservation range.
We consider two scenarios for the prior distribution: (i) uniform prior, $V_c/V_{\rm tot} \sim U(0,1)$, and (ii) a prior preferring local charge conservation, modeled by a Gaussian distribution in $V_c/V_{\rm tot}$ with the
mean value of $0.20$ and standard deviation of $0.05$~\footnote{These values are motivated by the analysis of net-proton cumulants in Ref.~\cite{Vovchenko:2024pvk}.},
$V_c/V_{\rm tot} \sim \mathcal{N}(0.20,0.05^2)$.

\begin{figure}[t!]
    \centering
    \includegraphics[width=0.49\textwidth]{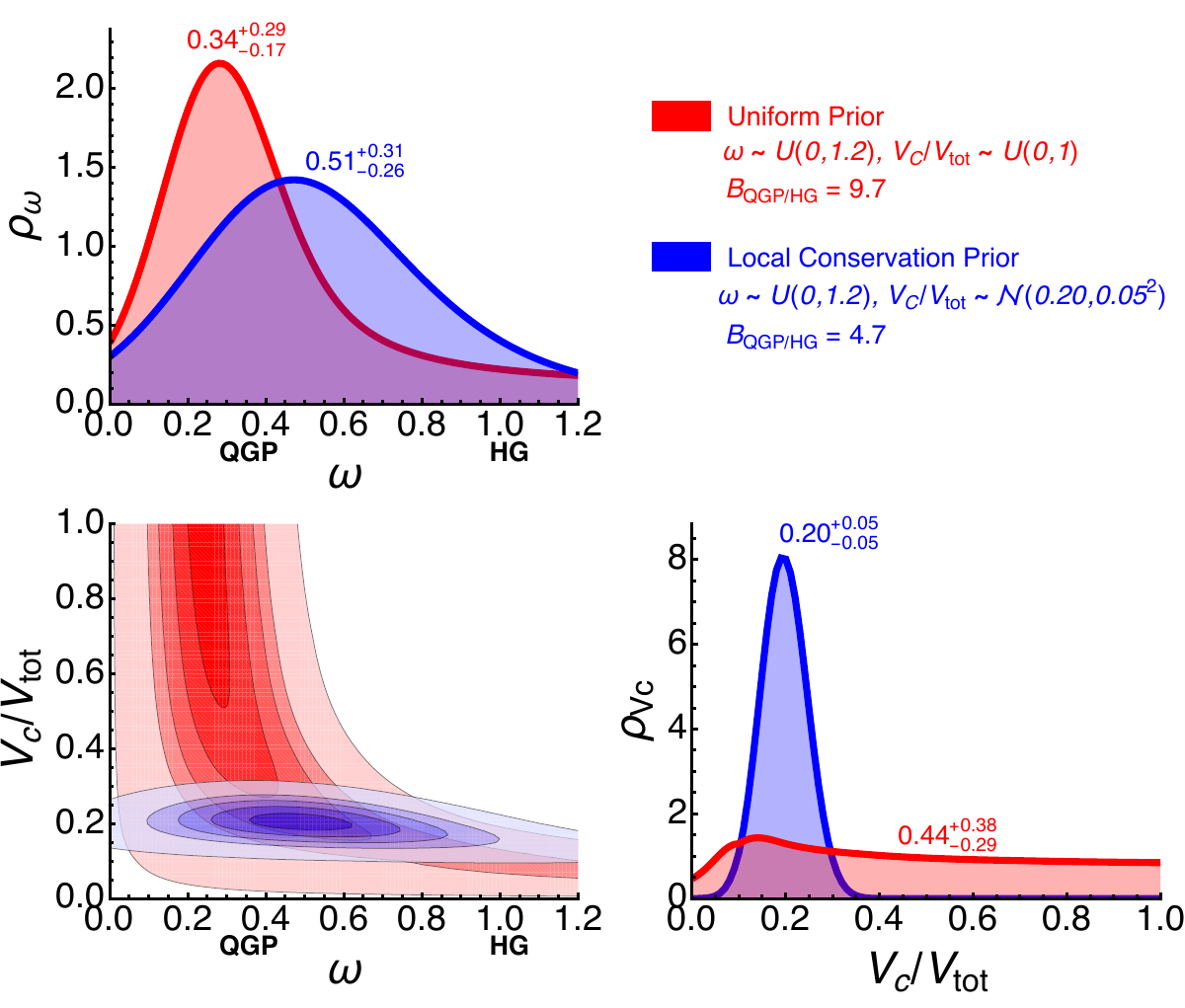}
    \caption{ 
    Posterior distribution for $\omega$ and $V_c/V_{\rm tot}$ obtained from our Bayesian analysis of $\sNN = 2.76$~TeV data employing uniform~(red) and local conservation~(blue) priors.
    The labels indicate the 68\% confidence interval around the median.
    The contours in the lower left plot represent 10\%, 30\%, 50\%, 68\%, and 95\% confidence regions for the joint $(\omega,V_c/V_{\rm tot})$ distribution.
    }
    \label{fig:DPlotAll}
\end{figure}

The posterior distributions are shown in Fig.~\ref{fig:DPlotAll}. 
For both scenarios, one observes only minor changes in the posterior distribution of $V_c$
in comparison to the prior, indicating that the D-measure is not very sensitive to $V_c$. 
One exception would be if the prior distribution for $\omega$ was strongly biased toward the HG scenario, as in this case, only a highly local $V_c$ could reasonably describe the data.
The posterior distribution for $\omega$ is more interesting.
For a uniform $V_c$ prior, one observes a considerable preference for QGP, with $\omega^{68\%}_{\rm CI} = 0.34^{+0.28}_{-0.16}$.
Contrasting QGP~($\omega_{\rm QGP} = 0.36$) and HG~($\omega_{\rm HG} = 1.1$) scenarios via the Bayes factor, one obtains $B_{\rm QGP/HG} = 9.7$ indicating moderate evidence~(bordering strong evidence) according to Jeffreys' classification scale~\cite{jeffreys1961theory,Lee_Wagenmakers_2014}. 
Giving a strong prior preference to a local charge conservation leads to $\omega^{68\%}_{\rm CI} = 0.53^{+0.30}_{-0.27}$ and $B_{\rm QGP/HG} = 4.7$ which still indicates moderate~(but close to anecdotal) evidence for QGP.

\emph{Predictions for 5.02~TeV.---}
The upcoming high-statistics measurements at $\sNN = 5.02$~TeV from the ALICE and CMS experiments~\cite{CMS:2023drv} during LHC Run 2 are expected to have notably reduced error bars and, thus, greater constraining power.
We performed calculations for the HG and QGP scenarios in central Pb–Pb collisions at $\sNN = 5.02$~TeV within our framework. 
The difference relative to the $\sNN = 2.76$~TeV case comes from a larger spatial rapidity coverage~($\eta_{\rm max} = 5.1$) and slightly modified blast-wave parameters~\cite{ALICE:2019hno}.
The results are qualitatively the same as those shown in Fig.~\ref{fig:DPlotRun1} and available in the supplemental material~\cite{SM}.
We note that a sizable portion of the experimental error at $\sNN = 2.76$~TeV comes from the difference between the $D'$ and $D''$ corrections.
Given that our calculations already incorporate local charge conservation, we advocate for a direct analysis of the uncorrected D-measure in Run 2 to 
enhance the discriminating power.

\emph{Summary and outlook.---}
We developed a new formalism for analyzing net-charge fluctuations in heavy-ion collisions, which allowed us to calculate the D-measure with the simultaneous treatment of the suppression of fluctuations due to QGP, correlations from resonance decays, local charge conservation, collective flow, and momentum acceptance.
We find that the hadron gas scenario is in fair agreement with the ALICE data for $\sNN = 2.76$~TeV Pb–Pb collisions only when a very short rapidity range of local charge conservation is enforced, while the QGP scenario is in excellent agreement with experimental data and largely insensitive to the range of local charge conservation.
A Bayesian data analysis reveals moderate evidence for the freeze-out of charge fluctuations in the QGP phase. 
The two scenarios can be tested further with the upcoming high-precision data from LHC Run 2.
The formalism 
offers ample opportunity for extensions, including the application to finite baryon densities and RHIC energies,  hadronic phase dynamics, variation of priors,  the analysis of high-order fluctuations, and balance functions.

\begin{acknowledgments}

\emph{Acknowledgments.} 
V.V. thanks Mesut Arslandok for the fruitful discussions.
This work was supported by the U.S. Department of Energy, 
Office of Science, Office of Nuclear Physics, under contract numbers DE-SC0022023 (J.P., C.R., V.V.), DE-SC0026065 (V.V.) and  DE-AC02-05CH11231 (V.K.).
This material is based upon work supported by the National Science Foundation under grants No. PHY-2514763, PHY-2208724 and PHY-2116686, and within the framework of the MUSES collaboration, under Grant No. OAC-2103680.

\end{acknowledgments}

\emph{Data availability.}
The data that support the findings of this article are openly available~\footnote{The data underlying the figures are available in a tabulated form on Zenodo at \href{https://doi.org/10.5281/zenodo.17613452}{10.5281/zenodo.17613452}}.

\bibliography{main}

\begin{appendix}
\setcounter{equation}{0}
\setcounter{figure}{0}
\renewcommand{\theequation}{A\arabic{equation}}
\renewcommand{\thefigure}{A\arabic{figure}}
\makeatletter
\section{\large \bf End Matter}

\section{Derivation of D-measure}

\emph{Charge susceptibility and resonance decays.---}
By construction, the charge susceptibility at hadronization is $\chi_2^{Q} = \omega \mean{n_{\rm ch}^{\rm prim}}$ where $\mean{n_{\rm ch}^{\rm prim}}$ is the charged particle density before decays.
One can decompose $\chi_2^Q$ as follows:
\eq{\label{eq:}
\chi_2^{Q} = \mean{n_{\rm ch}^{\rm prim}} + \varphi_2^{Q, \rm prim}.
}
The first term -- $\mean{n_{\rm ch}^{\rm prim}}$ -- is the Skellam distribution baseline corresponding to the uncorrelated production of charged hadrons. 
This term is also referred to as self-correlation~\cite{Pratt:2012dz,Ling:2013ksb}.
The second term -- $\varphi_2^{Q, \rm prim}$ -- describes correlations among the produced charges at hadronization. Given that $\chi_2^{Q} = \omega \mean{n_{\rm ch}^{\rm prim}}$, one has
\eq{
\varphi_2^{Q, \rm prim} = (\omega - 1) \mean{n_{\rm ch}^{\rm prim}}.
}
A value of $\omega$ 
different from unity implies the presence of (anti-)correlations among charges at hadronization.
In particular, the QGP scenario~($\omega_{\rm QGP} \approx 0.36$) implies a significant anti-correlation among charged hadrons following the hadronization.

Resonance decays during the hadronic phase produce additional correlations among final charged particles.
The most prominent such example is the $\rho^0 \to \pi^+\pi^-$ decay, which introduces a correlation between positively and negatively charged pions.
We incorporate this effect in the following way.
As the electric charge is conserved in all resonance decays, the full charge susceptibility  $\chi_2^Q$ does not change.
However, decays increase the final charged particle multiplicity, $\mean{n_{\rm ch}} = \gamma_Q \mean{n_{\rm ch}^{\rm prim}}$, leading to a rebalancing between the contributions from the self-correlation and two-particle correlations to $\chi_2^Q$. 
The parameter $\gamma_Q = \mean{n_{\rm ch}}/\mean{n_{\rm ch}^{\rm prim}}$ has been introduced in Eq.~\eqref{eq:omega} above and we employ the thermal model estimate $\gamma_Q \simeq 1.67$ throughout.
Writing the final charged susceptibility as
\eq{
\chi_2^Q = \mean{n_{\rm ch}} + \varphi_2^{Q}
}
and using $\chi_2^Q = \omega \mean{n_{\rm ch}^{\rm \rm prim}}$ and  $\mean{n_{\rm ch}} = \gamma_Q \mean{n_{\rm ch}^{\rm prim}}$ we get:
\eq{\label{eq:varphifinal}
\varphi_2^{Q} = \left( \frac{\omega}{\gamma_Q} - 1 \right) \mean{n_{\rm ch}}.
}

Resonance decays~($\gamma_Q > 1$) thus introduce additional
anti-correlations among charges in the final state.

\emph{2-point correlator and exact charge conservation.---}
To account for collective flow, charge conservation, and kinematical cuts, we employ a differential 2-point correlator,
\eq{\label{C2Q}
\mathcal{C}^Q_2(\rr_1,\rr_2) \equiv \mean{\delta \rho_Q(\rr_1) \delta \rho_Q(\rr_2)},
}
where $\delta \rho_Q(\rr_i) = \rho_Q(\rr_i) - \mean{\rho_Q(\rr_i)}$ is the net-charge density fluctuation at coordinate $\rr_i$.
Integrating $\mathcal{C}^Q_2(\rr_1,\rr_2)$ over a particular subvolume produces the variance of net-charge distribution of particles emitted from that subvolume.
In a recent work~\cite{Vovchenko:2024pvk}, the 2-point correlator $\mathcal{C}^Q_2(\rr_1,\rr_2)$ was derived for a thermal system with local charge conservation:
\eq{
\mathcal{C}^Q_2(\rr_1,\rr_2) = \chi_2^Q \left[ \delta(\rr_1 - \rr_2) - \frac{\varkappa(\rr_1,\rr_2)}{V_{\rm tot}} \right].
}
Here $\chi_2^Q$ is the grand-canonical charge susceptibility, $V_{\rm tot} = \int d \rr$ is the total system volume, and $\varkappa(\rr_1,\rr_2)$ 
is the local charge conservation kernel.
This last quantity satisfies the sum rule $\int d \rr_{1(2)} \varkappa(\rr_1,\rr_2) = V_{\rm tot}$.
In global equilibrium~(global charge conservation only) $\varkappa(\rr_1,\rr_2) = 1$ while local charge conservation can be modeled by $\varkappa(\rr_1,\rr_2)$ which is peaked around $|\rr_1 - \rr_2| = 0$.

At LHC, we integrate over the transverse plane and consider only the longitudinal (spatial) rapidity coordinate, $\rr \to \eta$.
We assume that the net-charge susceptibility $\chi_2^Q$ is independent of $\eta$, but we do allow for a non-uniform distribution in $\eta$ of the system volume, characterized by volume rapidity density $\rho_V(\eta) = dV/d\eta (\eta)$.
In this case, the volume element reads $d \rr = \rho_V(\eta) d \eta$ and $d \rr_1 d \rr_2 \delta(\rr_1 - \rr_2) = d \eta_1 d \eta_2 \rho_V(\eta_1) \delta(\eta_1 - \eta_2)$.

Therefore, the two-point correlator in spatial rapidity reads
\eq{\label{eq:C2Qfull}
\mathcal{C}^Q_2(\eta_1,\eta_2)  & = \rho_V(\eta_1) \chi_2^Q \left[ \delta(\eta_1 - \eta_2) - \rho_V(\eta_2) \frac{\varkappa(\eta_1,\eta_2)}{V_{\rm tot}} \right] \nonumber\\
& = \rho_V(\eta_1) \mean{n_{\rm ch}} \delta(\eta_1 - \eta_2) \nonumber \\
& \quad + \rho_V(\eta_1) \left(\frac{\omega}{\gamma_Q} - 1\right) \mean{n_{\rm ch}} \delta(\eta_1 - \eta_2)  \nonumber\\
& \quad - \rho_V(\eta_1) \rho_V(\eta_2) \frac{\omega}{\gamma_Q} \mean{n_{\rm ch}} \frac{\varkappa(\eta_1,\eta_2)}{V_{\rm tot}}.
}
Here, the first term corresponds to self-correlation, the second term describes local two-particle correlations, such as due to $\rho^0$ decays or hadronization of QGP, and the last term corresponds to non-local two-particle correlations due to exact charge conservation.

For the volume rapidity density, we considered a uniform~(boost-invariant) profile, $\rho_V(\eta) \propto \frac{\theta(\eta_{\rm max} - |\eta|)}{2 \eta_{\rm max}}$ and a Gaussian volume distribution, $\rho_V(\eta) \propto \exp\left(-\frac{\eta^2}{2 \sigma_V^2} \right)$.
The parameters $\eta_{\rm max}$ and $\sigma_V$ are fixed through the measurements~\cite{ALICE:2013jfw,ALICE:2016fbt} of the rapidity density of charged multiplicity~\cite{Vovchenko:2020kwg}.
The value of $\sigma_V$ reflects the measured width of the rapidity density of charged multiplicity, which is $\sigma_V = 3.86$ in Pb-Pb collisions at $\sNN = 2.76$~TeV~\cite{ALICE:2013jfw}.
The value of $\eta_{\rm max}$ in the uniform (boost-invariant) volume model is fixed such that the total volume matches the one in the Gaussian scenario~\cite{Vovchenko:2020kwg}.
This gives $\eta_{\rm max} = \sqrt{\pi/2} \, \sigma_V \simeq 4.83$.
We performed calculations in both the Gaussian and uniform volume scenarios and found the differences between the two scenarios to be virtually negligible for the charge fluctuations measured at midrapidity. 
We therefore employ the uniform volume distribution throughout the study for simplicity.

\emph{Kinematic cuts.---}
To account for collective flow and experimental cuts in momentum, we employ the binomial model for acceptance probability, following Ref.~\cite{Vovchenko:2024pvk}. 
Specifically, we assume that a final charged particle with spatial rapidity $\eta$ has momentum acceptance probability $p(\eta)$ that is independent of the other particles.
The variance of net charge in momentum acceptance is then obtained by folding the integration of $\mathcal{C}_2^Q(\eta_1,\eta_2)$ over spatial rapidities with the binomial distribution at each $(\eta_1,\eta_2)$ pair.
The self-correlation term in the 2-point correlator $\mathcal{C}_2^Q$ in Eq.~\eqref{eq:C2Qfull} is diluted by a factor $p(\eta_1)$, while the other terms correspond to two-particle correlations and are diluted by a factor $p(\eta_1) p(\eta_2)$.
The 2-point correlator describing the correlations of \emph{accepted} charge densities thus reads
\eq{\label{eq:C2Qacc}
& \mathcal{C}^{Q^{\rm acc}}_2(\eta_1,\eta_2) 
 = p(\eta_1) \rho_V(\eta_1) \mean{n_{\rm ch}} \delta(\eta_1 - \eta_2) \nonumber \\
& \quad + [p(\eta_1)]^2 \rho_V(\eta_1) \left(\frac{\omega}{\gamma_Q} - 1\right) \mean{n_{\rm ch}} \delta(\eta_1 - \eta_2)  \nonumber\\
& \quad - p(\eta_1) p(\eta_2) \rho_V(\eta_1) \rho_V(\eta_2) \frac{\omega}{\gamma_Q} \mean{n_{\rm ch}} \frac{\varkappa(\eta_1,\eta_2)}{V_{\rm tot}}.
}
Expression~\eqref{eq:C2Qacc} describes the correlation among charges emitted from spatial rapidities $\eta_1$ and $\eta_2$ which are subject to kinemtic cuts.
The total variance $\kappa_2[Q^{\rm acc}]$ measured in the experiment is obtained by intergrating $\mathcal{C}^{Q^{\rm acc}}_2(\eta_1,\eta_2)$ over all spatial coordinates,
leading to
\eq{\label{eq:k2Qacc}
\kappa_2[Q^{\rm acc}] & = V_{\rm tot} \mean{n_{\rm ch}}
\left[ \mean{p(\eta)} - \left(1 - \frac{\omega}{\gamma_Q}\right) \mean{p^2(\eta)}\right. \\
&- \left.\frac{\omega}{\gamma_Q} \mean{p(\eta_1) p(\eta_2)}_{\varkappa} \right].
}
Here,
\eq{\label{mean-p}
\mean{p(\eta)} & = \frac{1}{V_{\rm tot}} \int d \eta \, \rho_V(\eta) \, p(\eta), \\
\mean{p^2(\eta)} & = \frac{1}{V_{\rm tot}} \int d \eta \, \rho_V(\eta) \, [p(\eta)]^2, \\
\label{mean-p1p2}
\mean{p(\eta_1) p(\eta_2)}_\varkappa & = 
\frac{1}{V_{\rm tot}^2} 
\int d \eta_1 \rho_V(\eta_1) p(\eta_1) \nonumber \\
& \quad \times \int d \eta_2 \rho_V(\eta_2) p(\eta_2)  \varkappa(\eta_1,\eta_2)
}
are the acceptance probabilities averaged over spatial rapidities.
The mean acceptance probability $\mean{p(\eta)}$ also determines the mean charged multiplicity inside the acceptance, $\mean{N_{\rm ch}^{\rm acc}} = V_{\rm tot} \mean{n_{\rm ch}} \mean{p(\eta)}$.
Therefore, the D-measure inside momentum acceptance, $D = 4 \kappa_2[Q^{\rm acc}] / \mean{N_{\rm ch}^{\rm acc}}$, reads
\eq{\label{eq:Dmeasure}
D = 4 \left\{ 1 - \left( 1- \frac{\omega}{\gamma_Q} \right) \frac{\mean{p^2(\eta)}}{\mean{p(\eta)}} - \frac{\omega}{\gamma_Q} \frac{\mean{p(\eta_1) p(\eta_2)}_\varkappa}{\mean{p(\eta)}} \right\}.
}

\end{appendix}

\begin{appendix}
\widetext 
\setcounter{equation}{0}
\setcounter{figure}{0}
\renewcommand{\theequation}{S.\arabic{equation}}
\renewcommand{\thefigure}{S.\arabic{figure}}
\makeatletter
\section{\large \bf Supplemental material}

\section{Blast-wave model}

We calculate the charge acceptance probability $p(\eta)$ emitted from spatial rapidity $\eta$ as a weighted sum of pion, kaon, and proton contributions,
\eq{
p(\eta) = w_\pi p_{\pi}(\eta) + w_K p_K(\eta) + w_p p_p(\eta)
}
where the weights 
$w_\pi = 0.84$, $w_K = 0.12$, and $w_p = 0.04$
correspond to the relative midrapidity yields $dN/dy$ for pions, kaons, and protons measured by the ALICE Collaboration in 0-5\% Pb–Pb collisions at $\sNN = 2.76$~TeV~\cite{ALICE:2013mez}.
For $\sNN = 5.02$~TeV, the corresponding values are $w_\pi = 0.83$, $w_K = 0.13$, and $w_p = 0.04$~\cite{ALICE:2019hno}.

To calculate the acceptance probability $p_i(\eta)$ for a single particle species we use the blast-wave model~\cite{Schnedermann:1993ws} which incorporates the presence of longitudinal and radial collective flow.
The momentum distribution of a single particle emitted from spatial rapidity $\eta$ within the blast-wave model is given by
\eq{\label{eq:BW1}
\frac{d^3N_i}{p_T dp_T dy} (\eta) & \propto m_T \cosh (y - \eta) \, \int_0^1 \zeta \, d \zeta~{\rm exp}\left[-\frac{m_T \, \cosh \rho \, \cosh (y - \eta)}{T_{\rm kin}}\right] I_0 \left(\frac{p_T \sinh \rho}{T_{\rm kin}}\right),
}
where $\zeta\in [0,1]$ is the normalized radial coordinate in the transverse plane, $\rho = \tanh^{-1} ( \beta_s \zeta^{n_{\rm bw}})$ denotes the transverse flow rapidity, $y$ is the momentum rapidity, and $T_{\rm kin}$ is the kinetic freeze-out temperature of the system. The transverse mass is defined as $m_T = \sqrt{p_T^2 + m_i^2}$  with $m_{\pi}\approx138$~MeV, $m_{K}\approx494$~MeV, and $m_{p}\approx938$~MeV. The surface transverse velocity $\beta_s$ and the blast wave exponent $n_{\rm bw}$ control the maximum transverse velocity and the shape of the transverse flow velocity profile, respectively. 
We take the blast-wave parameters $T_{\rm kin} = 95$~MeV, $\beta_s = 0.883$, and $n_{\rm bw} = 0.712$  from~\cite{ALICE:2013mez} for 0-5\% Pb–Pb collisions at $\sNN = 2.76$~TeV and for 0-5\% Pb–Pb collisions at $\sNN = 5.02$~TeV we take $T_{\rm kin} = 90~{\rm MeV}$, $\beta_s = 0.906$, and $n_{\rm bw} = 0.735$ from~\cite{ALICE:2019hno}.

The D-measure measurements incorporate cuts in transverse momentum, 
$p_T^{\rm min} < p_T < p_T^{\rm max}$,  
and pseudorapidity, $|\tilde \eta| < \tilde \eta_{\rm cut}$.
One can map the distribution~\eqref{eq:BW1} from momentum rapidity $y$ to pseudorapidity $\tilde \eta$ through the variable change
\eq{
\sinh y = \frac{p_T}{m_T} \sinh\tilde\eta , \qquad
\qquad
dy = \frac{p_T \cosh\tilde\eta}{\sqrt{(p_T\cosh\tilde\eta)^2+m^2}} d \tilde\eta,
}
giving 
\eq{
\frac{d^3N_i}{dp_T d \tilde\eta}(\eta) = \frac{E(\tilde\eta,\eta) p_T^2 \cosh\tilde\eta}{\sqrt{(p_T\cosh\tilde\eta)^2+m^2}} \int_0^1 \zeta d\zeta~{\rm exp}\left[-\frac{E(\tilde\eta,\eta)~{\rm cosh}\rho}{T_{\rm kin}}\right] I_0 \left(\frac{p_T \sinh \rho}{T_{\rm kin}}\right)
}
where
\eq{
E(\tilde\eta,\eta) \equiv m_T~{\rm cosh}(y-\eta) = m_T~{\rm cosh}\left({\rm arcsinh}\left[\frac{p_T~{\rm sinh}\tilde\eta}{m_T}\right]-\eta \right).
}

The acceptance probability is calculated as
\eq{\label{eq:accept_prob}
p_i(\eta)  = \dfrac{\displaystyle \int_{p_T^{\rm min}}^{p_T^{\rm max}} dp_T \int_{-\tilde \eta_{\rm cut}}^{\tilde \eta_{\rm cut}} d \tilde \eta \frac{d^3N}{dp_T d \tilde\eta}(\eta)}{\displaystyle \int_{0}^{\infty} dp_T \int_{-\infty}^{\infty} d \tilde \eta \frac{d^3N}{dp_T d \tilde\eta}(\eta)}~.
} 
Here $p_T^{\rm min}=0.2$~GeV, $p_T^{\rm max}=5$~GeV and $0.1 < \tilde \eta_{\rm cut} < 0.8$ for 2.76~TeV data of ALICE~\cite{ALICE:2012xnj}.
We evaluate the integrals in~\eqref{eq:accept_prob} using Gauss-Kronrod quadrature for each value of $\eta$ and construct interpolating functions $p_i(\eta)$ for each acceptance cut under consideration.
We then use it to calculate the average acceptance probabilities $\mean{p(\eta)}$, $\mean{p^2(\eta)}$, and $\mean{p(\eta_1) p(\eta_2)}_{\varkappa}$.

\section{LHC Run 2 predictions}
\label{sec:run2}

\begin{figure*}[h!]
    \centering
    \includegraphics[width=0.48\linewidth]{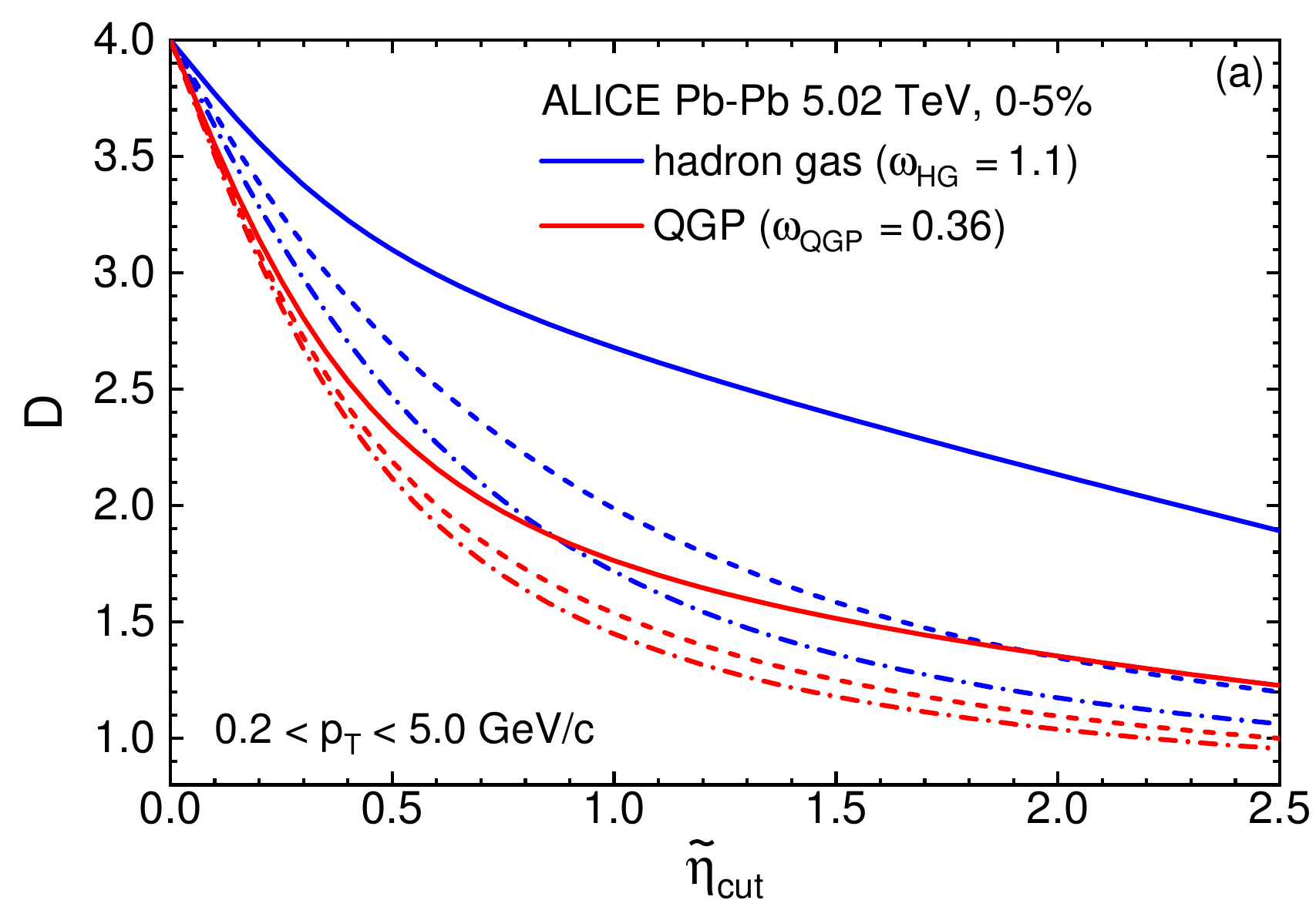}
    \includegraphics[width=0.48\linewidth]{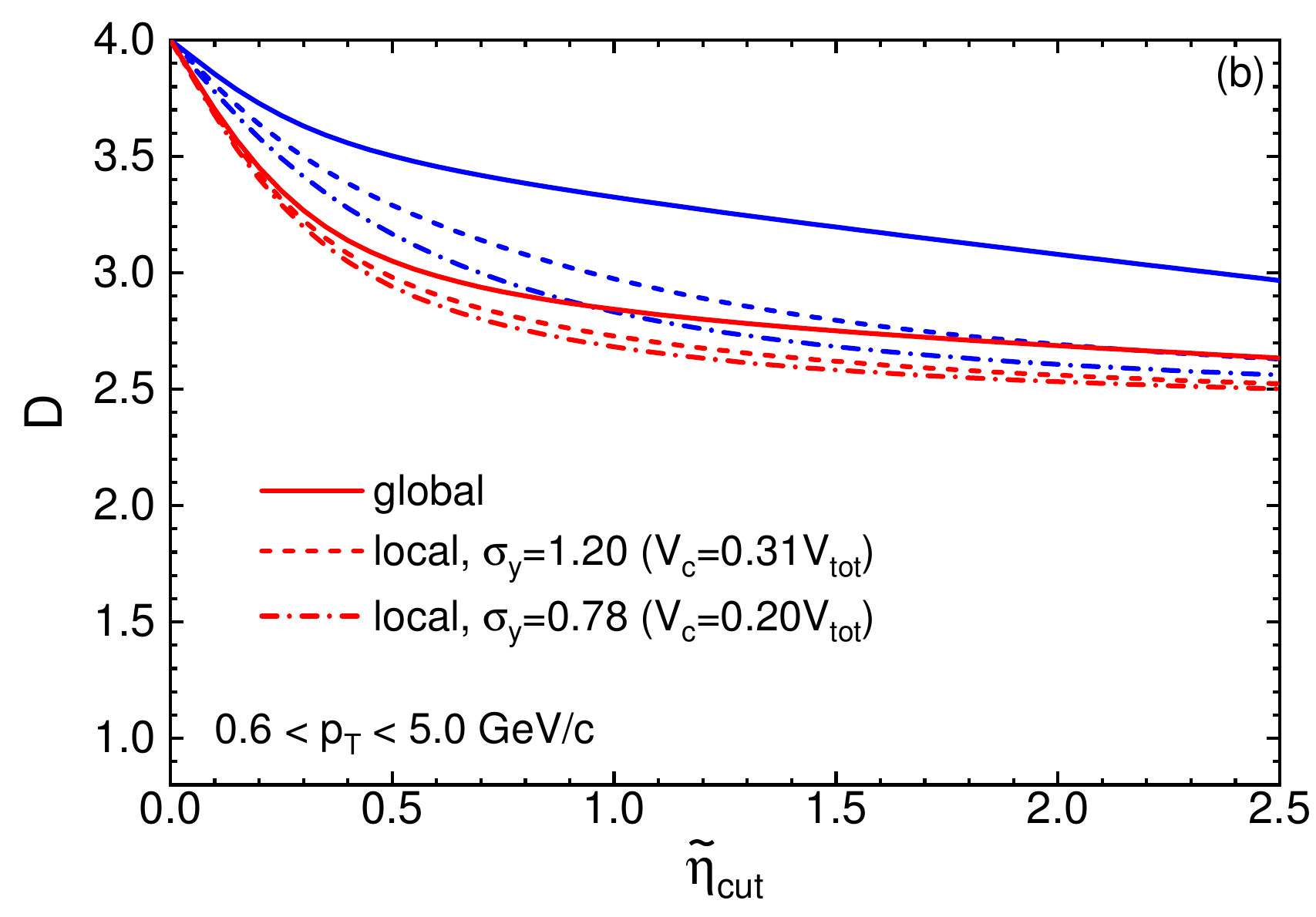}
    \includegraphics[width=0.48\linewidth]{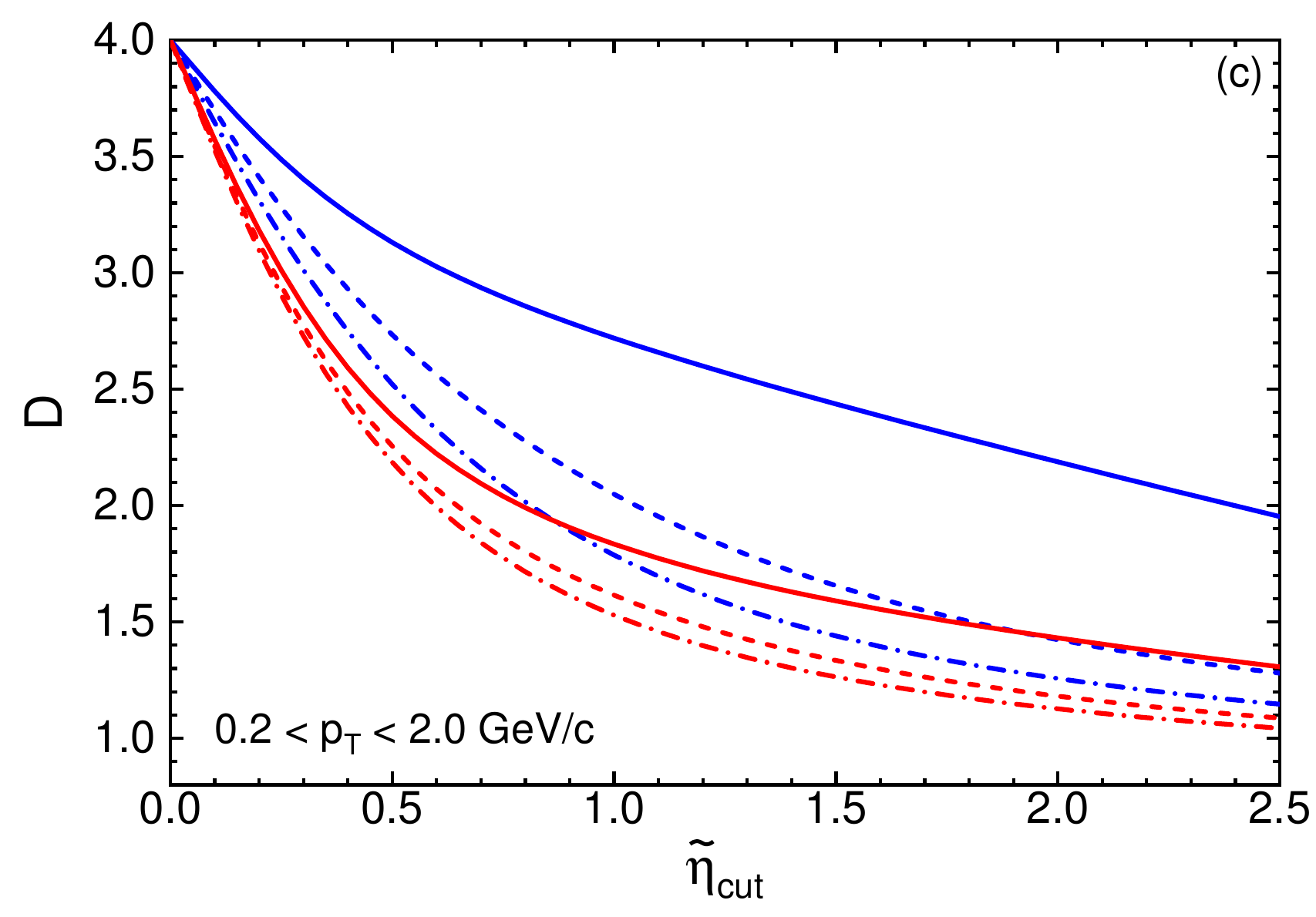}
    \includegraphics[width=0.48\linewidth]{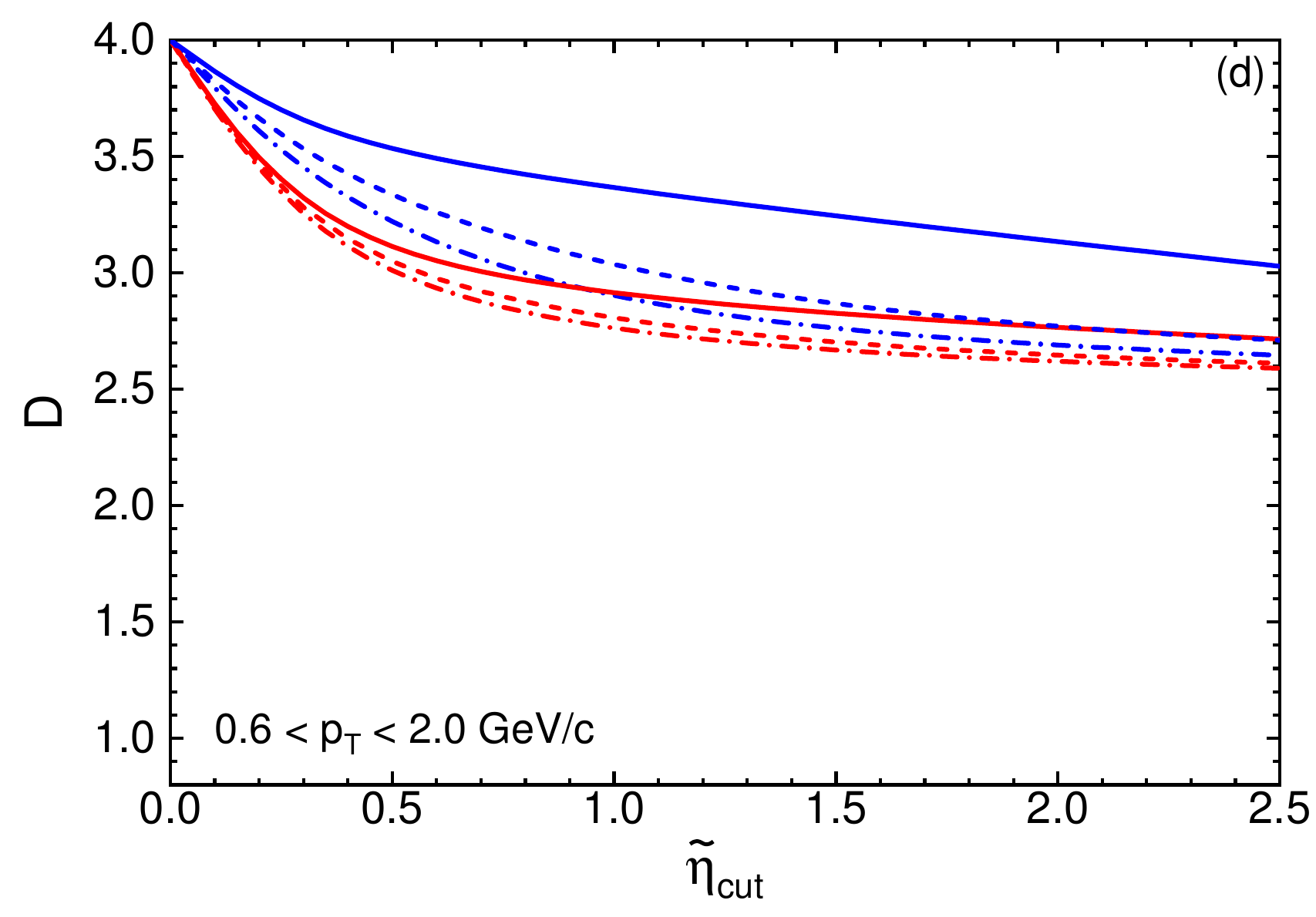}
    \caption{
    Predictions for the (uncorrected) D-measure in 0–5\% central Pb–Pb collisions at $\sNN = 5.02$~TeV, plotted as a function of the pseudorapidity acceptance $\tilde \eta_{\mathrm{cut}}$ for four different transverse momentum ($p_T$) ranges analyzed by the ALICE Collaboration. 
    The curves compare the hadron-gas (blue) and QGP (red) freeze-out scenarios under three assumptions about charge conservation: global (solid) and local with two different rapidity ranges, $\sigma_y = 1.20$~(dashed) and $\sigma_y = 0.78$~(dash-dotted). 
    }
    \label{fig:ALICERun2}
\end{figure*}

\begin{figure*}[h!]
    \centering
    \includegraphics[width=0.48\linewidth]{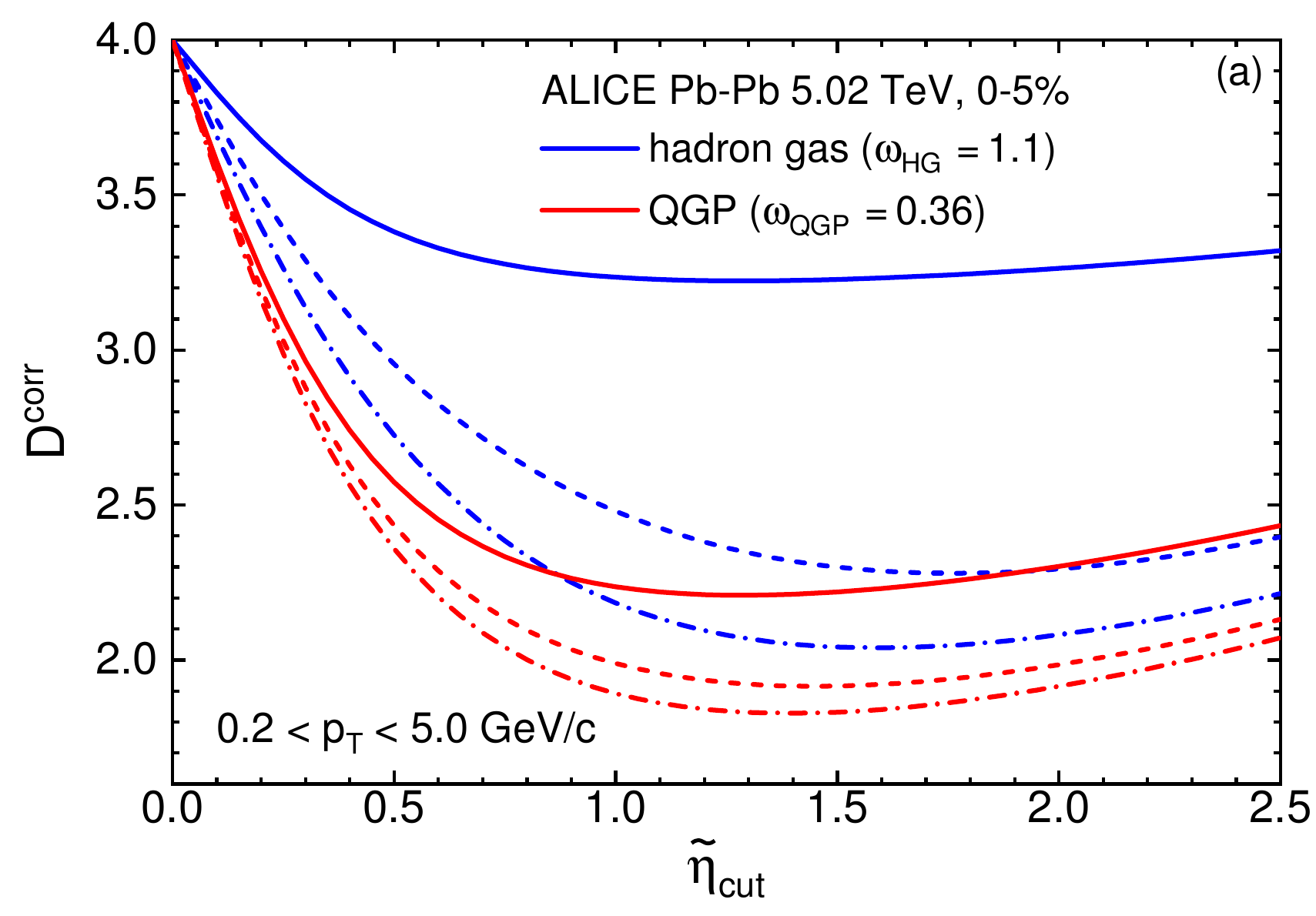}
    \includegraphics[width=0.48\linewidth]{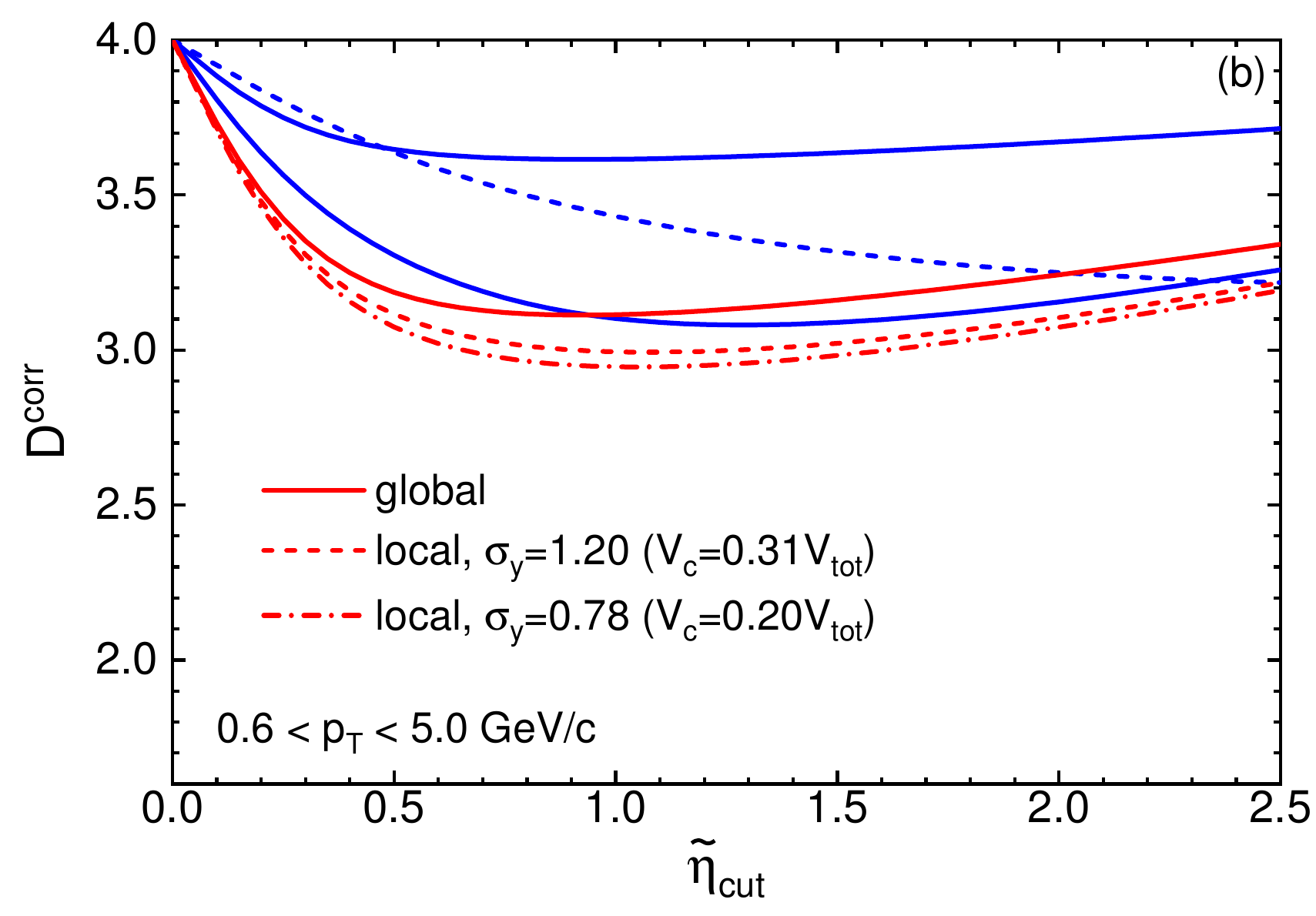}
    \includegraphics[width=0.48\linewidth]{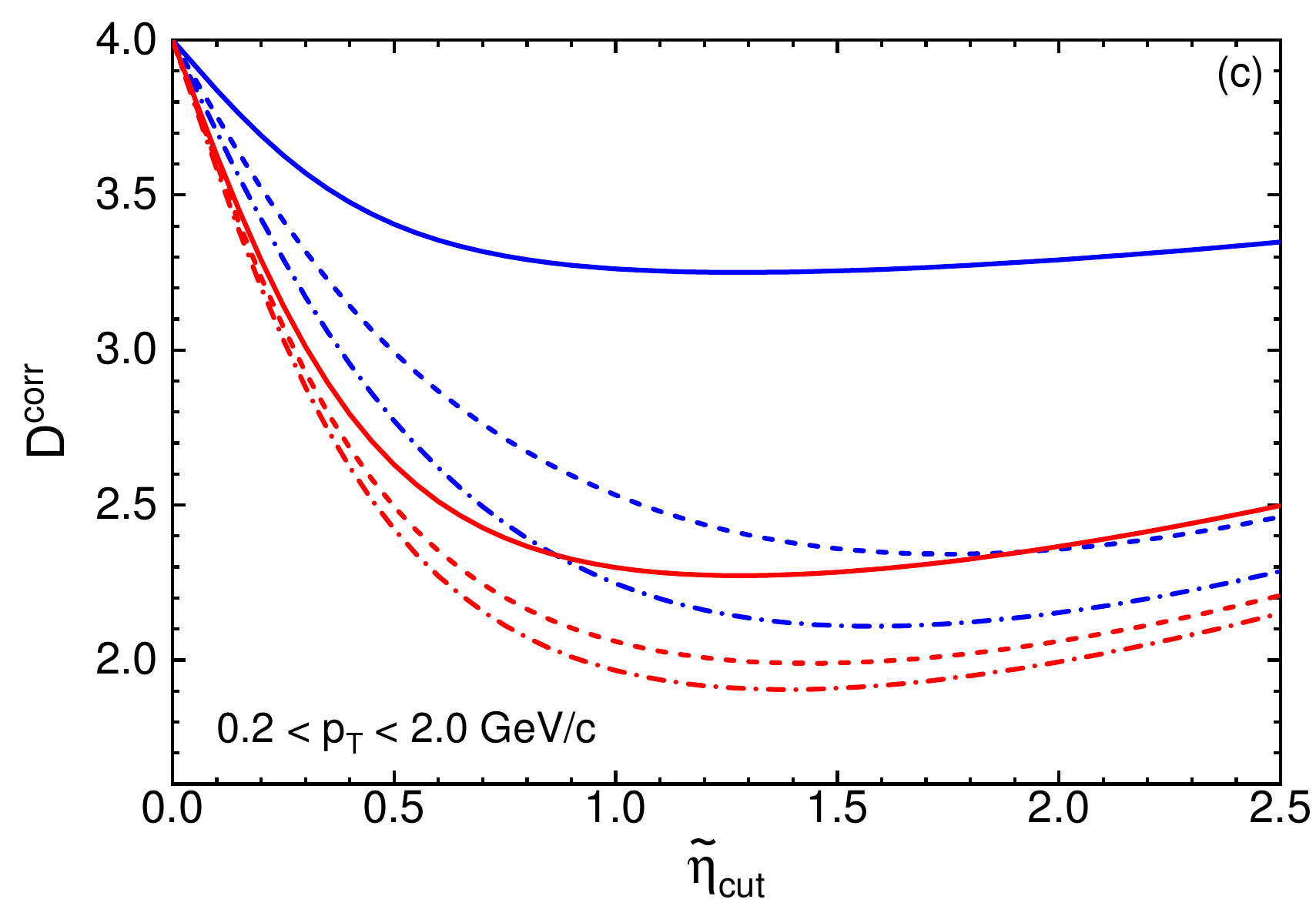}
    \includegraphics[width=0.48\linewidth]{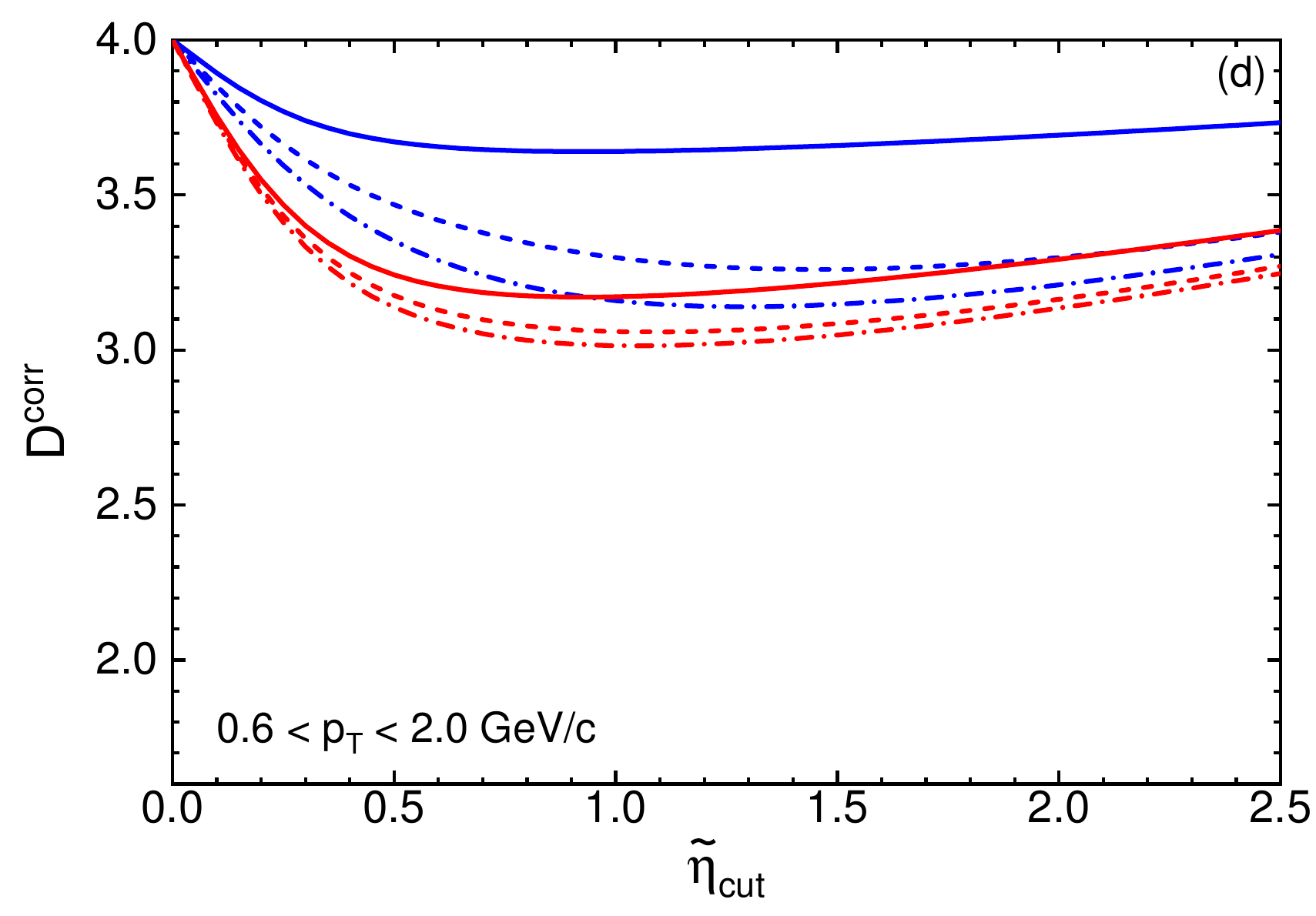}
    \caption{
    Same as Fig.~\ref{fig:ALICERun2} but for corrected $D$-measure ($D^{\rm corr}$).
    }
    \label{fig:ALICERun2-corr}
\end{figure*}

Here we present predictions for the D-measure in central Pb–Pb collisions at $\sNN = 5.02$~TeV from LHC Run 2.
The planned measurements by ALICE are performed in acceptance ranges $0.2(0.6) < p_T <2.0(5.0)$ GeV/$c$ at 0-5\% centrality as a function of pseudorapidity cut $\tilde \eta_{\rm cut}$.
Measurements at CMS are expected to extend acceptance coverage up to $\tilde \eta_{\rm cut}^{\rm max} = 2$~\cite{CMS:2023drv}.
Figure~\ref{fig:ALICERun2} depicts our predictions for the $\tilde \eta_{\rm cut}$ dependence of D-measure for the four $p_T$ ranges.
We present our predictions for the uncorrected D-measure, i.e. without the correction for charge conservation, which we argue to be most appropriate for direct comparisons with measurements.
One can see that the uncorrected D-measure is notably higher for calculations with $p_T^{\rm min} = 0.6$~GeV/$c$ compared to those with $p_T^{\rm min} = 0.2$~GeV/$c$.
This reflects stronger effects of charge conservation for a larger acceptance in $p_T$. We have also included predictions for the corrected D-measure (Figure~\ref{fig:ALICERun2-corr}) in case ALICE measurements include the correction factor.

\begin{figure}[h!]
    \centering
    \includegraphics[width=0.48\linewidth]{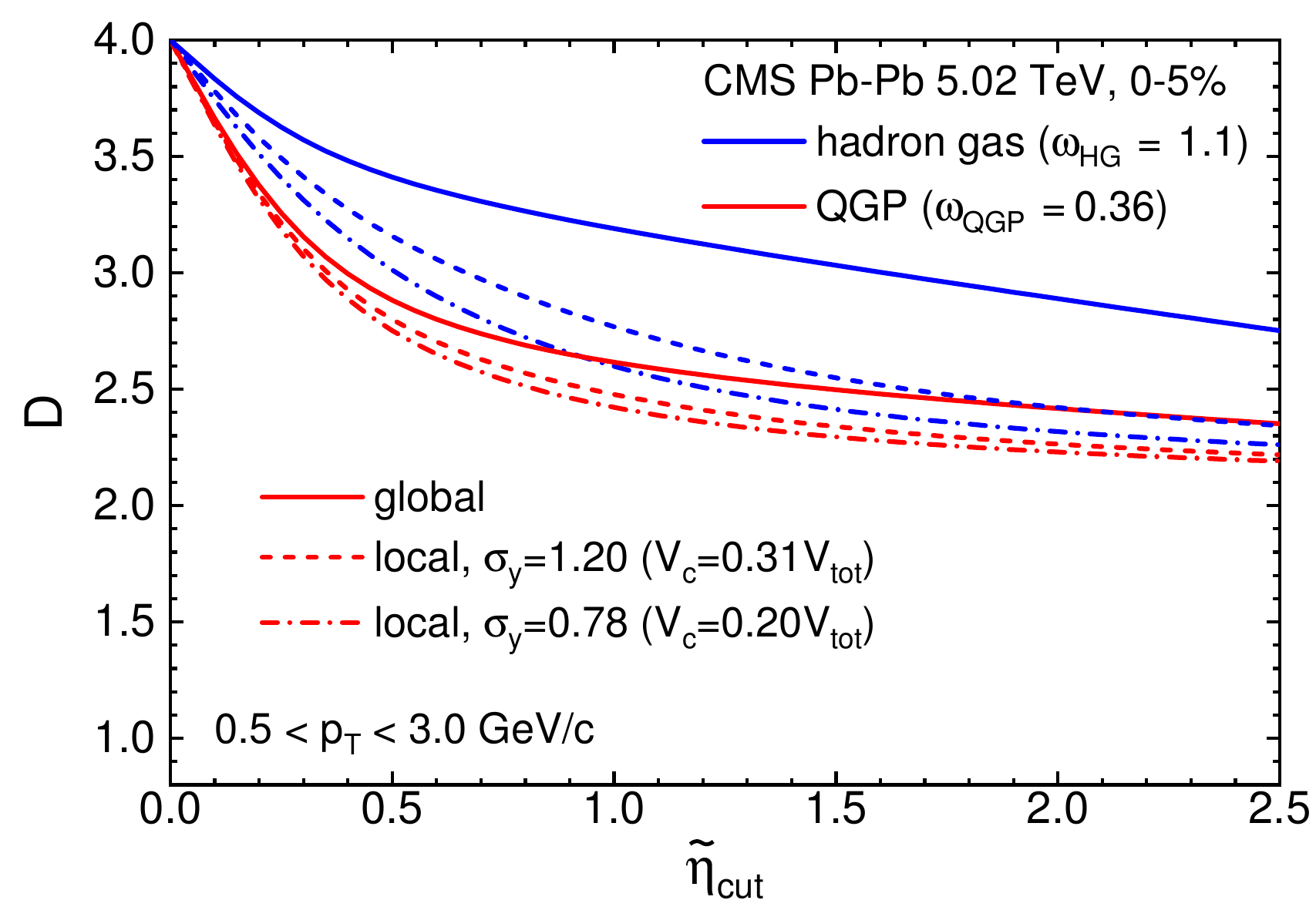}
    \caption{ Same as Fig.~\ref{fig:ALICERun2} but for transverse momentum range $0.5 < p_T < 3.0$~GeV/$c$ studied by the CMS Collaboration.
    }
    \label{fig:CMS}
\end{figure}

The planned measurements by CMS correspond to 0-5\% Pb–Pb collisions with transverse momentum cut $0.5 < p_T < 3.0$~GeV/$c$, as a function of pseudorapidity cut up to $\tilde \eta_{\rm cut} = 2.5$~\cite{CMS:2023drv}.
We present the corresponding predictions in Fig.~\ref{fig:CMS}.

\end{appendix}

\end{document}